\documentclass{PoS}
\usepackage{multirow}

\def\arcmin{\hbox{$^\prime$}}
\def\arcsec{\hbox{$^{\prime\prime}$}}
\usepackage{multicol}

\title{Status of ground based gamma-ray observations}

\ShortTitle{Rapporteur for gamma-ground for ICRC 2017}
\author{\speaker{Nahee Park}\\
        University of Chicago\\
        E-mail: \email{nahee@uchicago.edu}}
        
\abstract{This is a proceeding of a rapporteur talk given on ground-based gamma-ray astronomy at the $35^\mathrm{th}$ International Cosmic-Ray Conference (ICRC) held in 2017 in Busan, Republic of Korea. A total of $\sim$300 contributions were presented during the ICRC over 17 gamma-ray sessions. Here, I summarize the contributions mainly focusing on the source observations performed by ground-based gamma-ray instruments and the connection between gamma rays and cosmic rays. Any such summary must necessarily be incomplete. However, I have attempted to provide a glance into recent progress that has been made in using ground-based gamma-ray observations to understand the nature of high energy particles in our Universe.}

\FullConference{35th International Cosmic Ray Conference --- ICRC2017\\
		10--20 July, 2017\\
		Bexco, Busan, Korea}

\begin{document}
\section{Introduction}
Very high energy (VHE, E$>$100 GeV) gamma-ray observations reveal the extreme environments in our Universe where particles can be accelerated above TeV energies and allow us to study the dynamics of these high energy particles. We can learn how particles are accelerated in different classes of sources in various evolutionary stages and environments. This also opens a window to understand the origin and acceleration of high energy particles observed locally on Earth as cosmic rays. Since the first detection of VHE gamma-ray emission from the Crab Nebula in 1989 by the Whipple Observatory~\cite{1989Weekes}, VHE gamma-ray astronomy has been rapidly progressing. By the summer of 2017, the number of VHE gamma-ray sources has reached 198 with at least 8 broad source classes. The science topics studied by VHE gamma-ray observatories cover a wide range from gamma-ray astronomy to indirect dark matter detection, measurements of the local cosmic-ray fluxes and anisotropy, and particle physics. In this $35^{\textnormal{th}}$ International Cosmic-Ray Conference (ICRC), there were in total 17 gamma-ray sessions, which includes 289 contributions. Among these, about 225 contributions are closely related specifically to VHE gamma-ray observations. The considerable overlap with contributions from space-based gamma-ray observations highlights the scientific value of the multiwavelength approach that has become standard for many studies. Also, there are about 20 contributions from VHE gamma-ray observations to the other sessions, including the indirect cosmic-ray session and the dark matter session, and these contributions emphasize the multidisciplinary nature of the field. For this review, I will focus on the updates from the observations of gamma-ray sources in particular.

Since this is a review of the VHE gamma-ray studies presented during the cosmic-ray conference, it is worthwhile to point out the connection between cosmic rays and VHE gamma rays. VHE gamma rays are produced by the interaction of high-energy particles that will eventually become cosmic rays. Thus, understanding VHE gamma-ray sources is connected to understanding the sources of cosmic rays. Gamma-ray observations can provide what cosmic-ray observations cannot. Namely, because gamma rays are neutral particles their directional information is preserved and they can be used to find the source. This allows us to study sources that accelerate particles to high energies. We can study different source classes and the evolution of sources as accelerators. However, the gamma-ray studies have their own difficulties because the observed emission may come from different species of particles. Usually understanding the nature of the accelerator requires detailed models of the source region and evolution to disentangle leptonic and hadronic contributions. Also, unlike direct cosmic-ray measurements, hadronic gamma-ray emission is mostly dominated by protons and is not sensitive to different compositions. Compared to this, cosmic-ray observations generally cannot be used to detect the individual sources, and what we measure at Earth is likely contributed by multiple sources over millions of years. However, cosmic-ray measurements can provide detailed composition information, which allows us to study the source site composition, potential acceleration differences between elements, and propagation of the particles in our Galaxy by looking at the ratio between primary and secondary particles. We know that there should be high energy accelerators in our Universe to explain the remarkably smooth distribution of cosmic rays from several GeV up to $10^{21}$ eV. Answering what is the origin and acceleration mechanism for these high-energy particles is one of the most important tasks of VHE gamma-ray astronomy. Thus, the best way to understand the dynamics of high energy particles in our Universe is to combine the knowledge gained from cosmic-ray and gamma-ray observations.

\section{Instruments}
Ground-based gamma-ray observatories are designed to detect gamma rays by observing the air shower--the cascade of particles in the atmosphere. Because of the low flux from VHE gamma-ray sources, gamma-ray observatories have to overcome a high level of background events from cosmic-ray air showers. Largely there are two main techniques used to study the gamma-ray emission. One is observing the Cherenkov light emitted by secondary particles generated from the air shower. The other is observing the secondary particles themselves. Imaging Atmospheric Cherenkov Telescopes (IACTs) are designed to detect the Cherenkov light while air shower arrays are designed to detect the secondary particles. These two techniques provide complementary views of VHE gamma-ray sources.

In the IACT technique, the Cherenkov light from the shower arrives at ground level in a region with a radius of $\sim$100 m. This makes it easy to build a relatively small array of IACTs with a large effective area. Stereo measurements with multiple IACTs can improve the hadronic cosmic-ray rejection and thus the sensitivity. The current generation of IACT arrays consist of multiple telescopes, and these have the best instantaneous sensitivity to detect VHE gamma rays. A typical IACT detects VHE gamma-ray sources with 1$\%$ of the Crab Nebula flux in less than 30 hours. However, IACTs have a small duty cycle of $\sim$15$\%$ because the observation requires a clear, dark night. Also, their field of view (FoV) is relatively small ($<5^{\circ}$.)

Compared to the IACTs, the duty cycle of air shower arrays is $\sim$90$\%$ or larger and their FoVs represent a much larger area on the sky ($\sim$2 sr). With this large FoV, the air shower array can survey a large portion of the sky ($>$50$\%$) without pointing bias. 
The sensitivity and angular resolution of the air shower arrays are worse than IACTs, especially for energies lower than 1 TeV. The sensitivity of the current generation air shower array, the High Altitude Water Cherenkov (HAWC) observatory, is about ten times better compared to previous generation, and able to detect the Crab Nebula with 5$\sigma$ in one day. Because air shower arrays have a large FoV, they can also study very extended gamma-ray emission regions, such as emission from local accelerators or large Galactic structures, which are harder for IACTs to study. 

Here, I will briefly describe the ground-based gamma-ray observatories that provided major contributions to the scientific discussions during the conference.

\subsection{Imaging Atmospheric Cherenkov Telescopes}
The Very Energetic Radiation Imaging Telescope Array System (VERITAS) is an array of four IACTs located at the F. L. Whipple Observatory in southern Arizona, USA ($31^{\circ}40\arcmin$ N, $110^{\circ}57\arcmin$ W) at an altitude of 1,300 km above sea level (asl). Starting operation in 2007, VERITAS went through two major upgrades to improve the performance~\cite{Park2015}: the movement of one telescope in 2009, and upgrading the camera to high quantum efficiency PMTs in 2012. After these upgrades, VERITAS now observes gamma rays in the range between $\sim$85 GeV and $\sim$30 TeV. 
The FoV of VERITAS is about $3.5^{\circ}$ in diameter.

The Major Atmospheric Gamma-ray Imaging Cherenkov Telescopes (MAGIC) is an array of two IACTs with 17 m diameter reflectors, located on the Roque de los Muchachos Observatory, on the Canary Island of La Palma, Spain, at an altitude of 2,200 m asl. MAGIC started operation in 2004 with a single telescope. In fall 2009, a second telescope was added, improving the sensitivity and angular resolution. In 2012, another major upgrade of the system took place to make the stereoscopic system uniform~\cite{Aleksic2016}. The MAGIC telescopes were built to achieve the lowest possible energy threshold. With stereoscopic mode observation, the energy threshold is as low as 50 GeV. 
The FoV of MAGIC telescope is $3^{\circ}$ in diameter.

The High Energy Stereoscopic System (H.E.S.S.) is an array of five IACTs located in the Khomas Highlands of Namibia ($23^{\circ}16\arcmin18\arcsec$ S, $16^{\circ}30\arcmin01\arcsec$ E) at an altitude of 1,800 m asl. H.E.S.S. started operation in 2004 with an array of four IACTs, each with a 13 m diameter reflector. In 2012, a fifth telescope with a large 28 m reflector was added in the center of the array. Thanks to its large mirror area, it can be triggered by gamma rays with energies as low as 30 GeV. In 2016, H.E.S.S. completed an upgrade of the cameras for the four 13 m telescopes to reduce the dead time, reduce the failure rate, and to operate smoothly with the 28 m telescope which has higher trigger rates~\cite{Bonnefy805}. 
The field of view of H.E.S.S. is $5^{\circ}$ in diameter. 

The First G-APD Cherenkov Telescope (FACT) is a single IACT located at the Observatorio del Roque de los Muchachos on the Canary Island of La Palma, Spain. Observing since 2011, the main goal of FACT is to monitor the VHE gamma-ray activity of sources in the northern hemisphere. Because FACT's optical detectors are silicon photomultipliers (SiPMs) that maintain good performance under relatively bright light, FACT can continuously operate during bright moonlight. Thus, they have larger duty cycle than the other IACTs. FACT can detect a point source with a flux of $\sim$14$\%$ of the Crab Nebula flux within 50 hours of observation and its FoV is $4.5^{\circ}$~\cite{NoethePoS791_GA193} in diameter.   

\subsection{Air shower arrays}
The HAWC observatory is an air shower array with 300 water Cherenkov detectors located at an altitude of 4,100 m on Sierrra Negra, in the state of Puebla, Mexico ($18^{\circ}59\arcmin49\arcsec$ N, $97^{\circ}18\arcmin27\arcsec$ W.) The construction and installation of the full array was completed in 2015. HAWC monitors 2/3 of sky with an instantaneous FoV of $\sim$2 sr and with a duty cycle of 90$\%$. The one-year survey sensitivity of HAWC is $\sim$5--10$\%$ of the flux of the Crab Nebula~\cite{Abeysekara2017_ApJ843}.

\section{Galactic science}
\begin{figure}[ht!]
  \centering
  \includegraphics[scale=0.5]{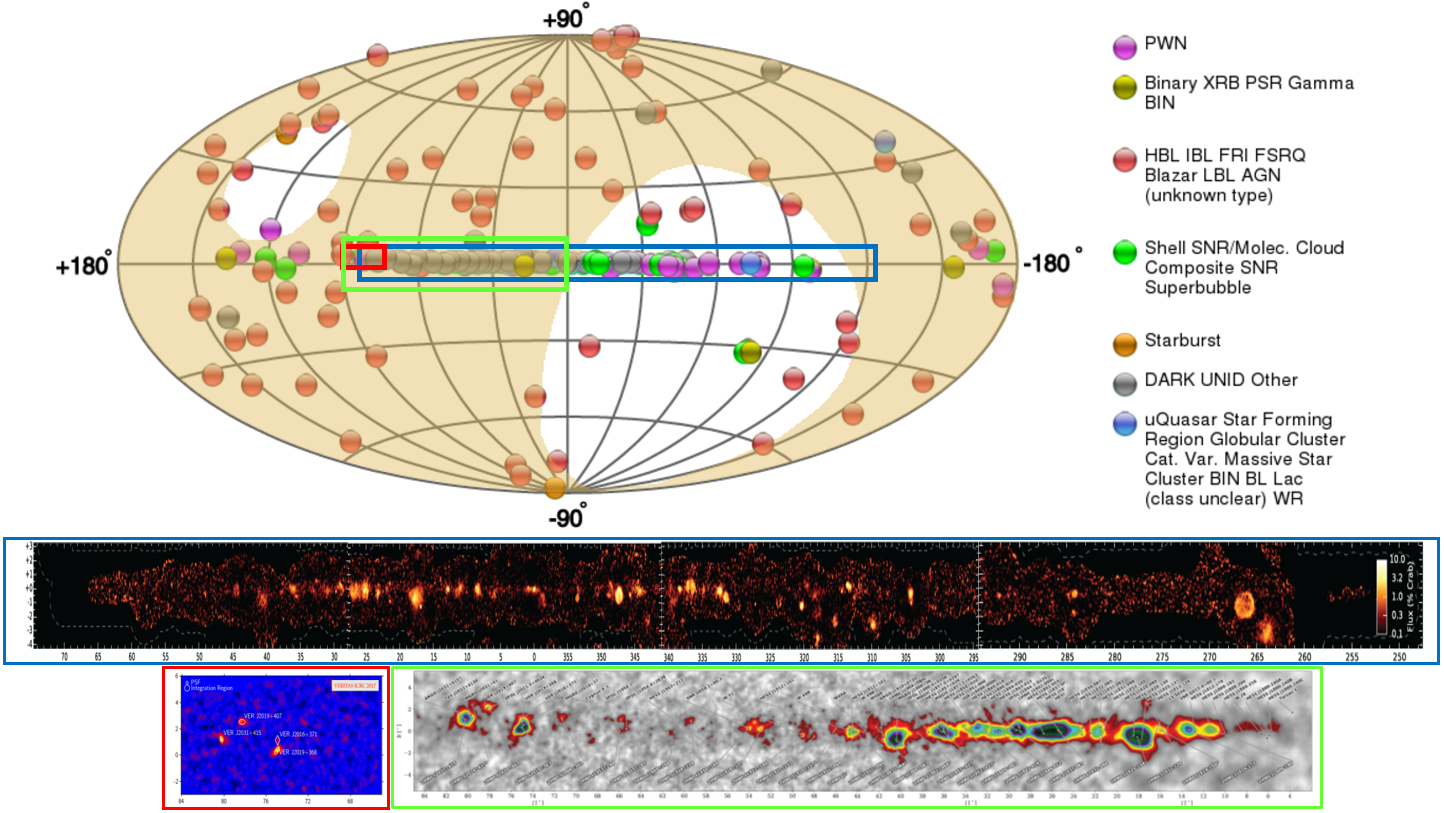}
\caption{VHE gamma-ray sources in Galactic coordinates as of summer 2017~\cite{tevcat}. The blue rectangular area indicates the region of the H.E.S.S. Galactic plane survey~\cite{Parsons697_GA232}. The red rectangular area indicates the region of VERITAS's Cygnus sky survey~\cite{Bird_GA187}. The green rectangular area indicates the Galactic plane shown by the HAWC collaboration~\cite{Abeysekara2017_ApJ843_40}. The yellow shaded area indicates the sky coverage of the HAWC observatory. The skymaps of each survey indicated with rectangular boxes are shown under the source map.} 
\label{Fig:GalacticSurvey}
\end{figure}
As shown in Figure~\ref{Fig:GalacticSurvey}, a large number of VHE gamma-ray sources are located in the Galactic plane. Due to the complicated environment and large density of possible sources, there are many unidentified VHE gamma-ray sources in our Galaxy. Classes of stellar objects identified to be the VHE gamma-ray emitters in our Galaxy include supernova remnants (SNRs), Pulsar wind nebulae (PWNs), binaries, and pulsars. 

While pointed observations can provide a detailed view of a single source, a survey of the Galactic plane with a uniform exposure provides an unbiased sample of the VHE gamma-ray source population in our Galaxy. This is a challenging project for IACTs due to their small FoVs. However, over ten years of observations, now we have the most sensitive Galactic plane survey data. H.E.S.S. surveyed the Galactic plane for almost 3,000 hours in the region shown in the blue rectangle in Figure~\ref{Fig:GalacticSurvey}. The survey identified a total of 78 sources including 12 PWNs, 8 composite objects, 8 SNRs, 3 binaries, 11 unassociated sources, and 36 sources that are not firmly identified~\cite{Parsons697_GA232}. VERITAS surveyed a $15^{\circ}$ by $5^{\circ}$ portion of the Cygnus region, shown as a red rectangle in Figure~\ref{Fig:GalacticSurvey}. The analysis of a total of 309 hours of observation, including survey, targeted, and follow-up observations, showed 4--5 sources. The difference in the number of VHE gamma-ray sources found in the H.E.S.S. Galactic plane survey and VERITAS's Cygnus survey are similar to the difference in the number of GeV gamma-ray sources reported by \textit{Fermi}-LAT (Large Area Telescope) in these two survey regions~\cite{Bird_GA187}.

The yellow shaded area in Figure~\ref{Fig:GalacticSurvey} shows HAWC's sky coverage. 
Recently, HAWC reported the detection of 39 VHE gamma-ray sources based on 507 days of observations~\cite{Abeysekara2017_ApJ843_40}. Among these, 19 sources were newly detected VHE gamma-ray sources. The Galactic plane observed by HAWC is shown in the lower right side of Figure~\ref{Fig:GalacticSurvey} surrounded by the green rectangular box. 

\subsection{Unidentified Sources}
The source and emission mechanisms of unidentified sources are unclear, but this category covers larger than half of the entire Galactic population. Searching for the nature of these sources presents challenges to both current and future experiments.

The VHE gamma-ray sources newly detected by HAWC presented a new set of unidentified sources, adding to 50 sources that are classified as either unassociated or not firmly identified from H.E.S.S.'s Galactic survey and VERITAS's Cygnus survey. Follow-up studies of these new HAWC sources were reported by IACTs during the conference. Among these 19 new sources, VERITAS reported the detection of VHE gamma-ray emission coincident with 2HWC J1953+294 and non-detections for 12 sources~\cite{Park696_GA241}. H.E.S.S. reported a non-detection for 2HWC J1928+177~\cite{Lopez-Coto732_GA70}. Because the IACTs' angular resolution is better than HAWC and their sensitivities are larger at lower energies, the non-detections of some of these sources provide information about the source extensions and/or spectral changes from a single power-law.

One of the most interesting unidentified sources was detected by \textit{Fermi}-LAT in the Cygnus superbubble~\cite{Ackermann2011}. Hard emission in a large cocoon area (best fit with a Gaussian kernel with a sigma of 2$^{\circ}$) requires high fluxes of recently accelerated high-energy particles to explain. Because of its large extension, this emission was mostly studied at VHE by the air shower arrays. During the conference, HAWC presented their study of this region with an improved sensitivity and angular resolution compared to previous studies~\cite{Hona710_GA243}. HAWC sees two emission regions--one point-like region from the nearby SNR $\gamma$--Cygni and one extended region around TeV J2032+4130, another not firmly identified source which is potentially a TeV PWN. The flux of the extended emission measured by HAWC is larger than the flux measurement of TeV J2032+4130 by IACTs, and matches with previous measurements of other air shower arrays. The extrapolated flux of HAWC's spectrum to the energy range of 1--100GeV agrees with the flux of the cocoon area measured by \textit{Fermi}-LAT. However, disentangling the emission from TeV J2032+4130 and a contribution from the cocoon emission may need further study. 

\subsection{Supernova Remnants}
Supernova Remnants (SNRs) have been suggested to be the source of Galactic cosmic rays up to the knee region ($\sim$3 PeV). One of the main goals of VHE gamma-ray observations is testing this hypothesis and studying SNRs to understand their evolution as the accelerators of cosmic rays in different environments. A handful of VHE gamma-ray emitting SNRs have been detected. However, disentangling the gamma-ray emission from leptonic particles and hadronic particles turns out to be a challenging task. It requires detailed models and knowledge of the surrounding area and evolutionary history of the source. 

Up to now, we have not identified a PeVatron, an accelerator that can accelerate particles up to PeV, among known SNRs. Young SNRs, including historic SNRs, have been suggested to accelerate particles to high energies. In this conference, there were at least two spectral cut-off measurements reported for young SNRs. The MAGIC collaboration reported that the VHE gamma-ray SED observed from a historic SNR, Cassiopeia A, prefers a power-law with exponential cut-off with a cut-off energy of 3.5 TeV~\cite{Guberman724}. H.E.S.S.'s observation of a young SNR, Vela Jr., shows that this source also prefers a power-law with exponential cut-off with a cutoff energy of 6.7 TeV~\cite{KominGA082}. As of now, all of the young SNRs' maximum energy observed in gamma rays is lower than 100 TeV. 

While Cassiopeia A is a pointlike source to IACTs, there are larger SNRs for which current IACTs can resolve the morphology. These sources may allow us to study region differences in the surrounding medium and the acceleration. MAGIC presented a morphological study combining their observations with \textit{Fermi}-LAT data to study the $\gamma$-Cygni region from the GeV energy range up to the VHE energy range~\cite{Strzys685}. MAGIC detected an extended emission region with a similar radius to the radio remnant as well as strong local emission in the northeast region that was previously detected by VERITAS. The morphology of the gamma-ray emission seems to change from low energies to high energies. It may require further multiwavelength observations and modeling to study the origin of the morphologies of VHE gamma rays in connection to the acceleration of particles in different regions in this source.

\subsection{Pulsar Wind Nebulae}
In PWNs, a fast rotating neutron star remaining after a supernova explosion provides an energetic outflow. A PWN is a magnetic bubble of relativistic particles formed by the outflow of the pulsar. When the outflow is confined by the surrounding medium, a standing termination shock is formed that can accelerate particles. TeV gamma-ray emission from PWNs can be explained as the inverse Compton scattering of energetic electrons and positrons in the ambient photon fields. The largest portion of Galactic VHE gamma-ray sources is associated with PWNs. A recently published PWN population study by H.E.S.S.~\cite{Abdalla2017} as well as previous studies pointed out that TeV PWNs observed by the current generation of IACTs are often associated with young pulsars with high spin-down luminosities. The size and morphology of TeV PWNs vary depending on the evolutionary stage of the PWN and the environment. The pulsar motion, SNR reverse shock, and particle diffusion can all change the distribution of VHE gamma rays in PWNs. Energy-dependent morphology studies of TeV gamma-ray emission from PWNs allow us to look into the history of particle acceleration and transportation in these sources. 

H.E.S.S's update on HESS J1825-137 shows a detailed view of energy-dependent morphological changes~\cite{Mitchell707_GA188}. For energies higher than $\sim$200 GeV, the gamma-ray emission decreases as the distance from the pulsar increases, with 50$\%$ of the emission at a distance of 0.5$^{\circ}$. This extension changes with energy, and the emission region looks smaller for energies higher than 32 TeV. The location of the maximum gamma-ray intensity also shifts toward the pulsar. As their previous publication pointed out, this may be related to the cooling of high-energy electrons. On the other hand, the update on the X-ray and gamma-ray emission from the Vela X PWN shows a uniform electron spectrum from 0.3 pc up to > 5 pc away from the pulsar~\cite{Tibaldo719_GA134}. Based on the strength of known magnetic fields in the region, we expect the cooling time of electrons to be shorter (4 kyr) than the age of the system (20 kyr). An efficient particle acceleration and/or transportation process in the source may be required to explain the morphology and spectrum. 
\begin{figure}[t!]
  \centering
  \includegraphics[scale=0.5]{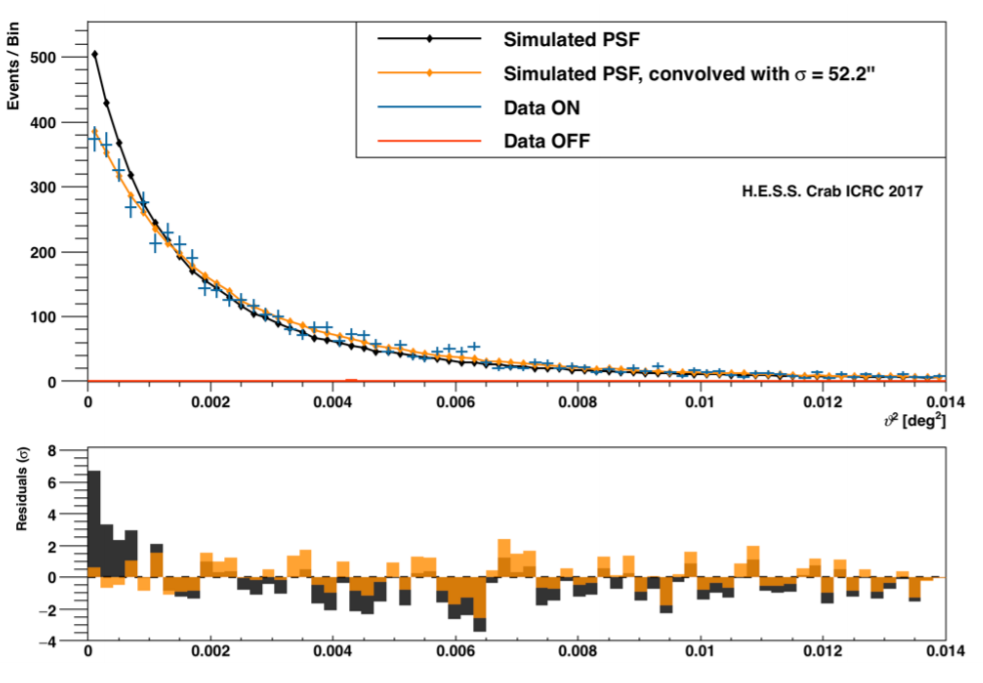}
\caption{$\theta^{2}$ histogram of gamma-ray-like events from the Crab Nebula (top). The blue line is the event distribution in ``On'' region and the red line shows the distribution in ``Off'' region. The bottom plot shows the residuals of the data with two different cases. The simulated PSF and the residuals of the data are shown in black. The simulated PSF convolved with the best-fit Gaussian distribution and the residuals of the data are shown in orange. } 
\label{Fig:CrabPWN_extension}
\end{figure}

A good angular resolution and good understanding of the instrument PSF are necessary to study these kinds of detailed morphological changes in the gamma-ray emission. Current IACTs typically quote about $0.1^{\circ}$ angular resolution, which is comparable to the angular extension of the Crab Nebula, the first detected TeV gamma-ray source. The Crab Nebula was a pointlike source in VHE gamma-ray observations. Upper limits on the extension were reported by IATCs. During this conference, H.E.S.S. reported the detection of extension of the Crab Nebula with a likelihood ratio test statistic of $\sim$83. As shown in Figure~\ref{Fig:CrabPWN_extension}, H.E.S.S. measures the best-fit 2D Gaussian sigma of $\sigma_\mathrm{2D,Gaussian}=52.2\arcsec \pm 2.9\arcsec_\mathrm{stat} \pm 2.9\arcsec_\mathrm{sys}$ for energies higher than 0.7 TeV~\cite{Holler676_GA307}. The extension seen in the gamma-ray emission is thus larger than the size of the X-ray nebula measured by \textit{Chandra}.

\begin{figure}[h!]
  \centering
  \includegraphics[scale=0.5]{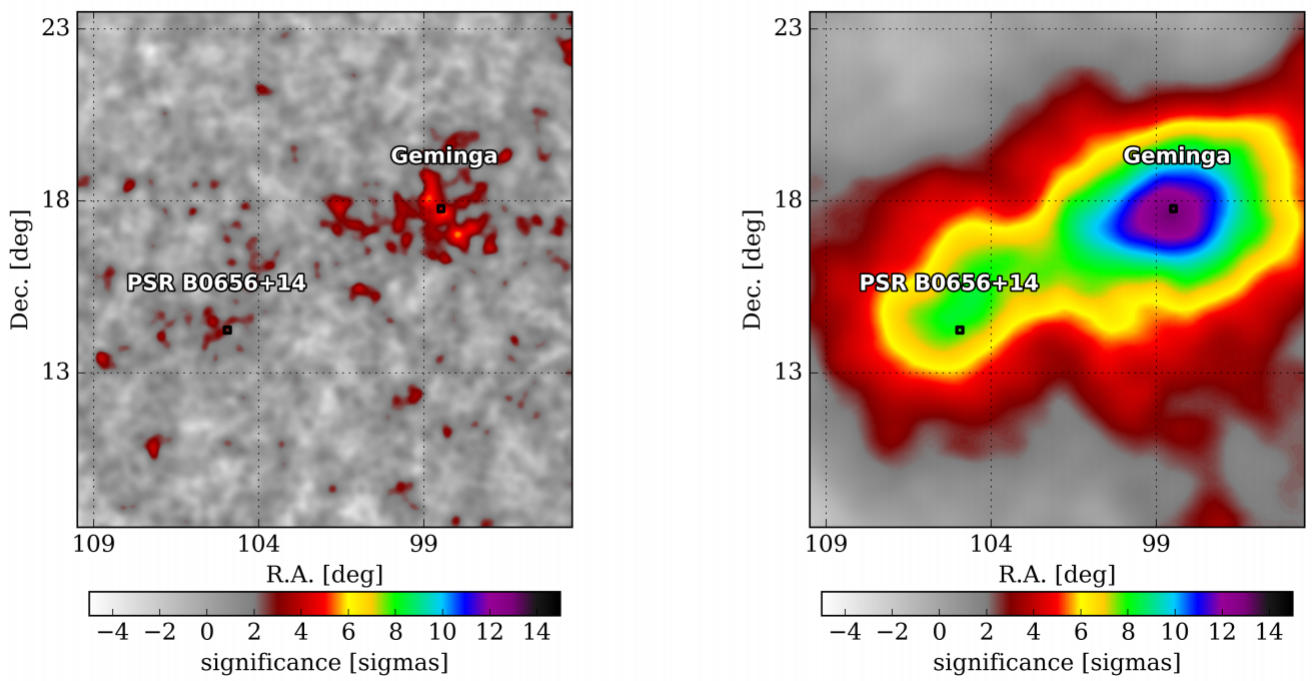}
\caption{The HAWC significance map for the Geminga and PSR B0656+14 region. The left side shows the skymap convolved with the point spread function and the right side shows the skymap convolved with the assumed particle diffusion from~\cite{Zhou690_GA250}. } 
\label{Fig:HAWC_localPWNs}
\end{figure}

Due to the size limitation of the FoVs of IACTs, the PWN population studied has been somewhat biased. With its wide FoV and improved sensitivity, HAWC can provide an unbiased sample of VHE gamma-ray emission from PWNs. Especially, HAWC is an ideal detector to detect very extended local PWNs. With 17 months of data, HAWC reported the detection of extended gamma-ray emission in the vicinity of Geminga and PSR B0656+14~\cite{Abeysekara2017_ApJ843_40,Greus722_GA122}. The skymaps of these sources are shown in Figure~\ref{Fig:HAWC_localPWNs}. These measurements are particularly interesting because both sources are very close to Earth (<300 pc). The electrons and positrons accelerated in these local accelerators may contribute to the local cosmic-ray spectra. Detailed studies of local accelerators will be important to find out the nature of the positron excess of cosmic-ray fluxes above 10 GeV measured by PAMELA, \textit{Fermi}-LAT, and AMS-02.

\subsection{Binaries}
\begin{figure}[t!]
  \centering
  \includegraphics[scale=0.5]{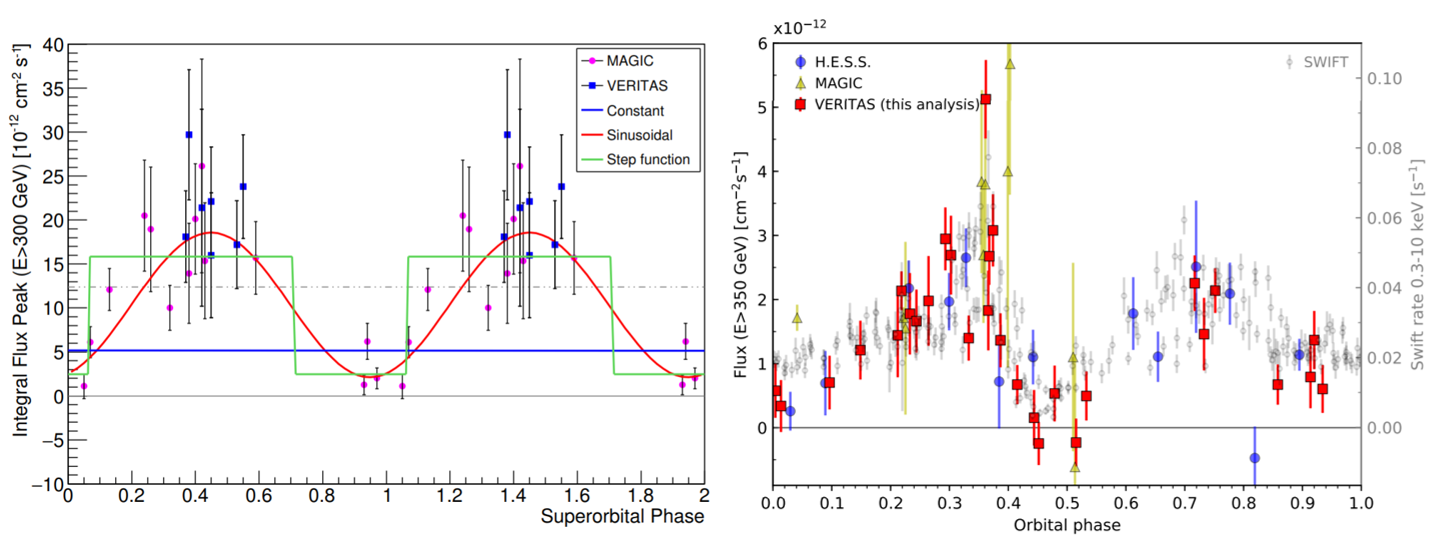}
\caption{VHE emission from LS I +61 303 (left) and HESS J0632+057 (right). The left plot shows VHE emission folded into the super-orbital phase in comparison to the super-orbital phase. The super-orbital phase is calculated based on radio measurements. Each data point represents the peak flux emitted in one orbital period during orbital phases 0.5 - 0.75. The results of fitting with three different functions are shown over the data (sinusoidal function, step function, and a constant flux). The gray dashed line shows 10$\%$ of the Crab Nebula flux. In the right panel, phase-folded VHE gamma-ray emission from HESS J0632+057 is shown. Colored markers show the VHE gamma-ray fluxes measured by VERITAS, MAGIC, and H.E.S.S. The gray markers denote X-ray observations with \textit{Swift}-XRT (0.3--10keV).} 
\label{Fig:Binaries}
\end{figure}

There are five binary systems known to emit VHE gamma-rays: PSR B1259-63, LS 5039, LS I +61 303, HESS J0632+057 and 1FGL J1018.6-5856. These systems are composed of a massive star and a compact object orbiting around the common center of mass. The compact object can be either a pulsar or a black hole. The identity of the compact object in these systems is unknown except in the case of the PSR B1259-63 system, where it is known to be a pulsar. Particle acceleration and gamma-ray emission mechanisms in these systems are not well known. Largely supported theories can be divided in two categories--the acceleration in an accretion-driven relativistic jet or in a shock interaction between the stellar wind and a pulsar wind. The periods of known VHE gamma-ray binaries vary from $\sim$4 days to $\sim$1237 days. The gamma-ray flux measured in the binary systems varies depending on the orbital phase. 
This variability is caused by changes in the environment for the particle acceleration and gamma-ray emission depending on the orbit phase. This makes the monitoring of the system in different phases to be very important. 
During this conference, the result of long-term monitoring campaigns on several binary systems showed interesting results. 

For LS I +61 303, VERITAS reported the detection of TeV emission around the entire orbit~\cite{Kar712_GA176}. Over ten years of observations from 2007 to 2017 show that there seems to be a baseline emission for the entire orbit with a flux of 3$\%$ Crab Nebula and a peak in the flux around apastron. MAGIC reported on studies of the super-orbital variability of gamma-ray emission from LS I +61 303 at orbital phases of 0.5-0.75 with a period of 1,667 days by folding the peak emission with the 1,667 days of super-orbital period measured in radio wavelengths~\cite{Hadasch725_GA116}. As shown on the left side of Figure~\ref{Fig:Binaries}, it is hard to describe the folded gamma-ray emission with constant emission. The current data set can be explained by either a sinusoidal function or a step function with a p-value of 7--8$\%$.

A clear correlation between X-ray and gamma-ray fluxes over the 315 day orbital period of HESS J0632+057 as shown in the right panel of Figure~\ref{Fig:Binaries} was reported by VERITAS~\cite{Maier729_GA78}. Measurements of both the X-ray and VHE gamma-ray bands show a high flux level around the phases of 0.2-0.4 and 0.6-0.8. All of the spectra of VHE gamma-ray measured in different phases can be described by power-law distributions. The spectral indexes measured in different phases have similar values, suggesting similar conditions for the gamma-ray emission.

H.E.S.S. represented the monitoring of PSR B1259-63/LS 2883 system over multiple years around its periastron~\cite{Romoli675}. VHE gamma rays were only detected around periastron for this binary with a 3.4 year long orbit. TeV SEDs and light curves measured in 2004, 2007, 2011, and 2014 generally agree with each other, indicating similar processes for VHE gamma-ray emission. The Combined light curve shows a double-peak structured light curve with a local minimum flux around the time of periastron. Brightening of the source at about 30 days before the periastron passage was also reported. 

The properties of the gamma-ray emission are not the same among the five binary systems that are confirmed to emit VHE gamma rays. By increasing the sample of TeV binary systems, our understanding of the acceleration of particles in different environmental conditions will be improved. There are continuous efforts to detect new VHE gamma-ray binary systems by multiple collaborations~\cite{Maier729_GA78, Hadasch725_GA116, Rho742_GA33}. In this conference, H.E.S.S. announced the detection of a sixth binary system, Eta Carinae, in our Galaxy, thanks to the lower energy threshold and an improved low-energy sensitivity provided by its large 28 m central telescope. This binary system is composed of a luminous blue variable and its lighter companion star, which makes it the first colliding-wind binary system detected in the VHE gamma-ray band. The period of its orbit is 5.5 years and the gamma-ray emission is detected during the last periastron passage in 2014. 

Among other binary candidates, a binary system containing PSR J2032+4127 was recently identified and has a long orbital period of 40--50 years. The system is located on the top of a bright TeV PWN candidate, TeV J2032+4130. 
The periastron of this system will be around November of 2017, providing a unique opportunity to study the system~\cite{Bird189_GA706}. It is expected for VHE gamma-ray emission to reach to the highest flux level between periastron and superior conjunction~\cite{Bednarek749}. Depending on the efficiency of the acceleration and the environmental conditions, the gamma-ray emission may overcome the emission from TeV J2032+4130.


\subsection{Other Galactic Sources}
There are other classes of VHE gamma-ray emitters in our Galaxy. One of the most interesting sources is the center of our Galaxy, where H.E.S.S. found evidence to support an accelerator that can produce PeV particles. In this conference, MAGIC reported no variability of VHE gamma-ray emission from the Galactic center based on a multi-year monitoring campaign between 2012 and 2015~\cite{Lopez721_GA124}. 

In the VHE band, the flux of Galactic diffuse emission is generally weaker than that of the sources because of its soft spectrum. H.E.S.S. reported the detection of diffuse emission in the TeV energy range, which may include fluxes of unresolved sources due to the limitation of the survey sensitivity. 
In this conference, HAWC presented their analysis status for detecting the Galactic diffuse emission, describing how they will account for the emission from unresolved sources~\cite{Zhou689_GA251}. 


\section{Sources in the Large Magellanic Cloud}
The Large Magellanic Cloud (LMC) is a dwarf satellite galaxy of our Galaxy. At a distance of 50 kpc, it is possible for current IACTs to search for individual accelerators in this Galaxy. 
At the last ICRC, H.E.S.S. reported the detection of three sources, PWN N 157B, SNR N 132D, and a superbubble 30 DorC~\cite{Komin2015}, in the LMC. This year, H.E.S.S. reported the detection of VHE gamma rays from a binary system in the LMC--P3~\cite{Komin730_GA76}. Significant gamma-ray emission above energies higher than $\sim$700 GeV was detected during an orbital phase between 0.2 and 0.4. Compared to this, the emission detected by \textit{Fermi}-LAT is peaked at an orbital phase of 0. Similar in luminosity to the PWN N 157B in the LMC, LMC P3 is the most luminous gamma-ray binary ever detected.

\section{Extragalactic Science}
There are 70 extragalactic sources detected in the VHE gamma-ray band. Except for two starburst galaxies, the rest of the extragalactic VHE gamma-ray emission is connected to the nature of active galactic nuclei (AGNs)--compact regions at the center of galaxies. The generally accepted hypothesis suggests that a supermassive black hole resides at the core of the AGN. As material falls into the black hole, gravitational energy is released and some of the energy can ultimately get converted into the kinetic energy of an outflow, forming well-collimated jets of plasma that move with relativistic speeds~\cite{Krawczynski2013}. Particles can be accelerated in jets either by magnetic reconnection or by internal shocks. The accelerated particles will emit broadband electromagnetic spectra. A typical non-thermal energy spectrum of VHE gamma rays emitted from AGNs shows two bumps. The low-energy bump corresponds to synchrotron radiation from relativistic electrons in the jet. Under the leptonic production scenario, the high-energy bump corresponds to inverse Compton scattering by the same electron population. The target photons are either the synchrotron photons themselves, referred to as the synchrotron self Compton (SSC) model, or an ambient photon field, referred to as an external Compton (EC) component. Hadronic production mechanisms for the high-energy bump have also been advanced, but these are more complicated than in the leptonic case.

VHE gamma-ray AGNs can be further classified as blazars and radio galaxies. Radio galaxies are AGNs with visible jet structures because the angle between the line of sight and the direction of their jets is large. Four radio galaxies are detected in VHE gamma-ray band. The rest, 64 VHE gamma-ray emitting AGNs, are blazars--AGNs with a jet pointing toward us. A more detailed review of gamma-ray emission in AGNs can be found in~\cite{Madejski2016}. Depending on the properties of their emission, blazars can be largely divided between BL Lacertae (BL Lac) objects and flat spectrum radio quasars (FSRQs). FSRQs usually show strong and broad emission lines and are typically more distant objects. Because of their distances and the lower energy at which their synchrotron emission peaks (around the infrared range), FSRQs are usually faint in the VHE gamma-ray band. Only six such objects have been detected as of this conference. BL Lac objects show weak or no emission lines. Depending on where their synchrotron emission peaks, these objects are further classified into low-frequency peaked BL Lacs (LBLs), intermediate-peaked BL Lacs (IBLs), and high-frequency peaked BL Lacs (HBLs). Up to now, 2 LBLs, 8 IBLs, and 47 HBLs were detected in the VHE gamma-ray band. 

During this ICRC, the detection of four more blazars was reported--OJ 287~\cite{OBrien650_GA80} and RGB J2056+496~\cite{Benbow641_GA108} by VERITAS, PKS 0736+017~\cite{Cerruti627_GA168} by H.E.S.S., and OT 081~\cite{Sitarek658_GA58} by MAGIC. OT 081 is an LBL detected in 2016 while its flux was elevated in GeV, X-ray, and optical bands. PKS 0736+017 is an FSRQ with a redshift of 0.189 detected during gamma-ray flaring in the MeV--GeV energy band in 2015. Night-to-night flux variation was reported as well as a study of the location of the gamma-ray emission region. RGB J2056+496 is one of the brightest, hard-spectrum objects in the \textit{Fermi}-LAT 2FHL catalog coincident with an X-ray and a micro-quasar candidate, LS III +49 13. OJ 287 was detected in 2017 during a historically high state observed in the X-ray band. This source shows 12 year periodic outbursts in the optical band, indicating that there may be a binary supermassive black hole system in its core.     

\subsection{Intergalactic Environment: Extragalactic Background Light}
\begin{figure} 
  \centering
  \includegraphics[scale=0.4]{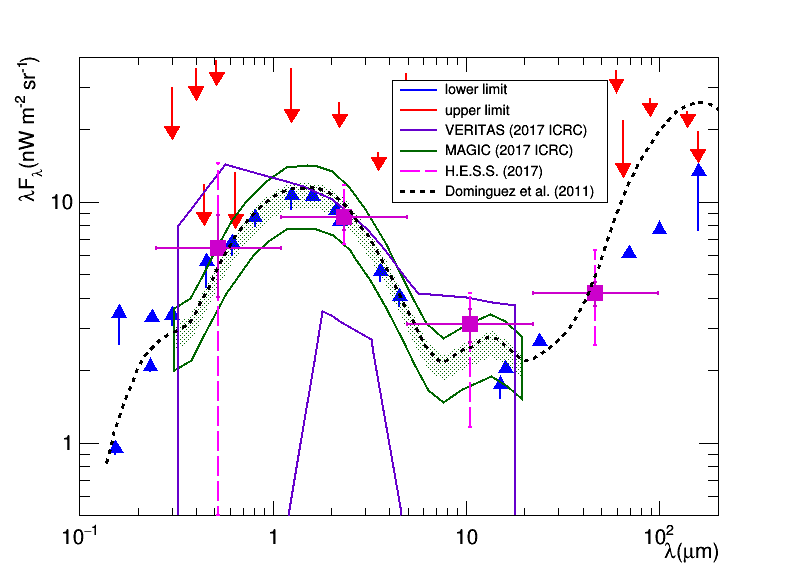}
\caption{The EBL SED measured by IACTs. Lower limits plotted with blue arrows are estimated from counting galaxies. Upper limits plotted with red arrows are estimated from the direct night sky background light. H.E.S.S.'s measurements~\cite{Abdalla2017b} are shown with magenta squares with a solid line showing the statistical errors and a dashed line showing the systematic errors. VERITAS's measurement~\cite{Pueschel1107} is shown with a dark violet contour. MAGIC's measurements~\cite{Olaziola604} are shown in green with the shaded area showing the statistical errors and the darker contour showing the systematic errors. MAGIC's result shown here uses Dominguez's model~\cite{Dominguez2011}} 
\label{Fig:EBL_2017}
\end{figure}
There is an intrinsic limitation on how far we can observe the Universe with VHE gamma rays because of the interaction of VHE gamma rays with the extragalactic background light (EBL), which is the integrated light emitted throughout the Universe. 
As of now, there are only two blazars detected in the VHE gamma-ray band above a redshift of 0.9--S3 0218+35 with a redshift of 0.944 and PKS 1441+25 with a redshift of 0.939. Because the absorption of VHE gamma rays softens the spectrum of the sources measured at Earth, understanding the EBL is essential to study the intrinsic spectrum for most blazars. On the other hand, the EBL can be measured with VHE gamma-ray observations by studying the level of flux suppression from distant VHE gamma-ray sources if the distance and the intrinsic spectrum of the source are known. 

The most up-to-date EBL measurements by major IACTs are shown in Figure~\ref{Fig:EBL_2017}. H.E.S.S. and VERITAS reported EBL measurements independent of any assumed spectral shape, whereas MAGIC reported measurements of the normalization for several specific EBL models. To perform these measurements, H.E.S.S. selected nine HBLs for their study to avoid the bias from the potential gamma-ray absorptions around the source regions~\cite{Abdalla2017b}. The study was done to estimate the shape of the EBL SED and the normalization by simultaneously fitting the intrinsic spectra together with a generic attenuation. The EBL signature is preferred at the 9.5$\sigma$ level compared to the null hypothesis. VERITAS estimated the effect of EBL absorption based on a generic EBL shape from random spline generation~\cite{Pueschel1107}. By repeating this procedure, they estimated the confidence intervals and upper limits. The preliminary results with eight HBLs are shown in Figure~\ref{Fig:EBL_2017} with violet solid line. More blazars will be added for future studies. 
MAGIC reported their updated EBL measurement with 12 blazars with MAGIC data alone and with MAGIC and \textit{Fermi}-LAT data combined~\cite{Olaziola604}. The analysis method used the shape of EBL SED from three different EBL models and estimated the best fitting normalization for the EBL. The results shown in Figure~\ref{Fig:EBL_2017} with the green shaded area and line are for MAGIC data alone with and assumption of EBL SED shape suggested by \cite{Dominguez2011}. 
It is interesting to see that the model independent studies by H.E.S.S. and VERITAS show a shape and intensity that are similar to the lower limit set by galaxy counts. Although the approaches are different, all three results agree well with each other and generally lie closer to the lower limit of EBL measurements rather than the upper limit.
Further details of the EBL measurements with VHE gamma-ray observations can be found in the highlight talk presented during the conference~\cite{Pueschel1107}.

\subsection{Monitoring of the Sources}
AGNs are strongly variable sources with episodes of remarkable flux increases. Observations of flaring sources allow us to study the jets of supermassive black holes and their environment, how and where particles are accelerated and emit non-thermal radiation, and eventually the energy dynamics around the supermassive black hole itself.
Because of the flux increase during the flare that can exceed the nominal flux by more than an order of magnitude, flares provide an opportunity to detect new sources as well. In fact, among four detections reported during this conference, three were caught during a flare. Measurements of new sources located far away can also provide good data to study EBL. 

Flare observations are often triggered by detections in other wavelengths, from radio up to the GeV energy range. In particular, GeV gamma-ray satellites such as \textit{Fermi}-LAT generally have a larger FoV and duty cycle than IACTs. Thus, they can provide semi-continuous observation of a large number of sources. Also, because of the proximity of their energy range to that of VHE gamma-ray observatories, they provide a good trigger for flaring activities in the VHE gamma-ray band. Still, the most direct way to detect VHE gamma-ray flaring from blazars is catching the flare with VHE gamma-ray observatories. Indeed, all major IACTs have carried out monitoring campaigns for the blazars although the coverage of their monitoring is limited due to their small FoV, and duty cycle. 

FACT and HAWC are relatively newly added VHE gamma-ray observatories and have provided wide coverage of the monitoring for extreme flares. 
FACT, a single IACT, has a higher energy threshold and lower sensitivity compared to larger arrays of IACTs. However, it can observe under bright moonlight without large degradation in its sensitivity because SiPMs, the optical sensors FACT uses for the focal plane, can operate under bright background light. Since its installation in 2011, FACT has been monitoring the activities of VHE gamma-ray sources especially for the brightest blazars including Mrk 421, Mrk 501, and 1ES 1959+650~\cite{Dorner609_GA238}. As an air shower array with a large FoV ($\sim$2 sr), HAWC has monitored a large number of sources simultaneously with a sensitivity that can detect the Crab Nebula with 5$\sigma$ significance per one day~\cite{Martinez-Castellanos656_GA62}. Up to now, HAWC reported the detection of two nearby blazars, Mrk 421 and Mrk 501, and their activities. Daily fluxes of these two objects were monitored by both FACT and HAWC. The comparison study between these two instruments shows a good correlation over $\sim$200 nights of observations with a few exceptions possibly due to fast variability of the sources~\cite{Dorner625}. 

\subsection{Notable flares}
Due to the intrinsic variability of blazars, the best way of studying the details of particle acceleration and gamma-ray emission during the flaring is to build models based on data that is taken simultaneously across the electromagnetic spectrum. Because of this, we have seen a growing number of fast communications between multiwavelength observatories. Thanks to this large community-wide effort, many reports we hear nowadays are based on a wide range of multiwavelength observations. 
Here, I will briefly mention a few notable flares that were discussed during the conference.

\begin{figure}[th!]
  \centering
  \includegraphics[scale=0.5]{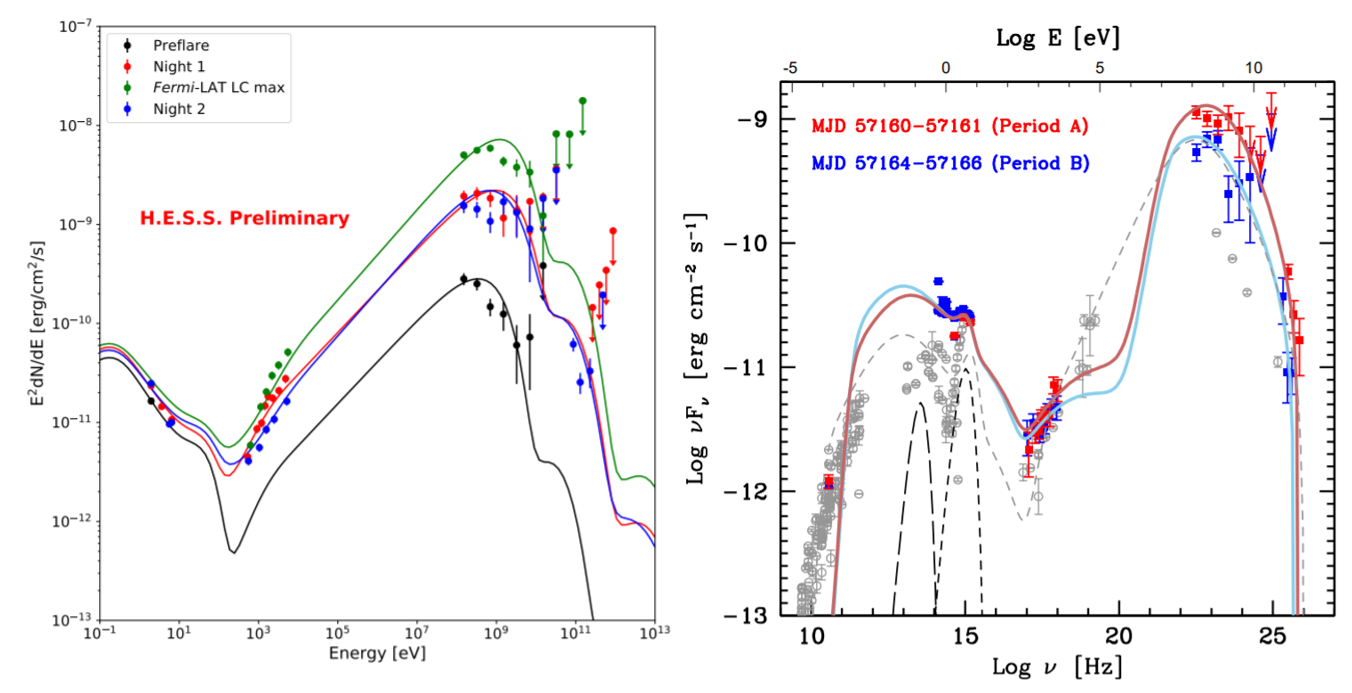}
\caption{SED of 3C279 during the 2015 flaring activity, shown with a hadronic emission model~\cite{Diltz2015} (left). On the right side, the SED of PKS 1510-089 during its flare in 2015 is shown with a one-zone external Compton model.} 
\label{Fig:FSRQ_flares_SEDs}
\end{figure}

All three FSRQ flares reported during the conference either show a detection or hint of an energy spectral break between the GeV gamma-ray emission measured by \textit{Fermi}-LAT and the VHE gamma-ray emission measured by IACTs, as shown in Figure~\ref{Fig:FSRQ_flares_SEDs}. If both the GeV emission and the VHE gamma-ray emission are produced by inverse Compton scattering of the same population of leptonic-dominant particles in the external photon field, the spectral break and flux difference can be interpreted as the result of absorption of VHE gamma rays in the dense photon field of the broad line region (BLR) of FSRQs. Based on these assumptions, the level of opacity and the location of gamma-ray emission zone can be calculated. H.E.S.S. reported the measurements of flares from PKS 0736+017 in 2015~\cite{Cerruti627_GA168}, 3C279 in 2016~\cite{Romoli649_GA81}, and PKS 1510+089 in 2016 (this also includes measurements from MAGIC)~\cite{Zacharias655_GA68} and estimations of the opacity and the location of emission region. The estimated opacity, defined as the log of the ratio of the flux extrapolated from the GeV energy range to the observed flux, varies from 2 to 5, and the distance between the emission region and the black hole is about $10^{17}$ cm. Among these, the flare of 3C279 in 2015 was recorded as a historical maximum flux in the GeV energy range for the source. With dedicated observation, \textit{Fermi}-LAT measured an intrinsic variability timescale of 2 minutes. Unlike the large variability observed in the gamma-ray energy range, no significant flux changes were measured in the lower energy bands where the synchrotron energy peak was measured in a broadband SED. These observations challenge the conventional leptonic model. A hadronic model~\cite{Diltz2015} shown on left side of Figure~\ref{Fig:FSRQ_flares_SEDs} may describe the data better.

Non-variability of the low-energy flux with large flux changes in the gamma-ray band was also observed during the 2015 flare activities in PKS 1510-089 by MAGIC~\cite{Sitarek657_GA60}. The SED with a one zone leptonic model with external emission could describe the data with different assumptions including one in which the emission zone is outside of the BLR. An extensive multiwavelength campaign was carried out during the flare, resulting in the measurement of the rotation of optical polarization and the observation of a new radio component that may be connected to the gamma-ray flares~\cite{Sitarek657_GA60}. Another strong flare from PKS 1510-089 was detected by H.E.S.S. and MAGIC in 2016, with a peak flux of $\sim$ 0.8 times the flux of the Crab Nebula above 200 GeV~\cite{Zacharias655_GA68}. A variation of VHE gamma-ray flux on time scales less than an hour was detected. While the VHE flux varied by more than a factor of 10 in a single night, only a modest flux variation of $\sim$30$\%$ was reported in the optical band. There was no significant flux variation in the GeV band. However, the spectral index of GeV band was significantly hardened, indicating that the peak of the inverse Compton component was shifted by more than a factor 10 to higher energies during the flaring.

\begin{figure}[t!]
  \centering
  \includegraphics[scale=0.6]{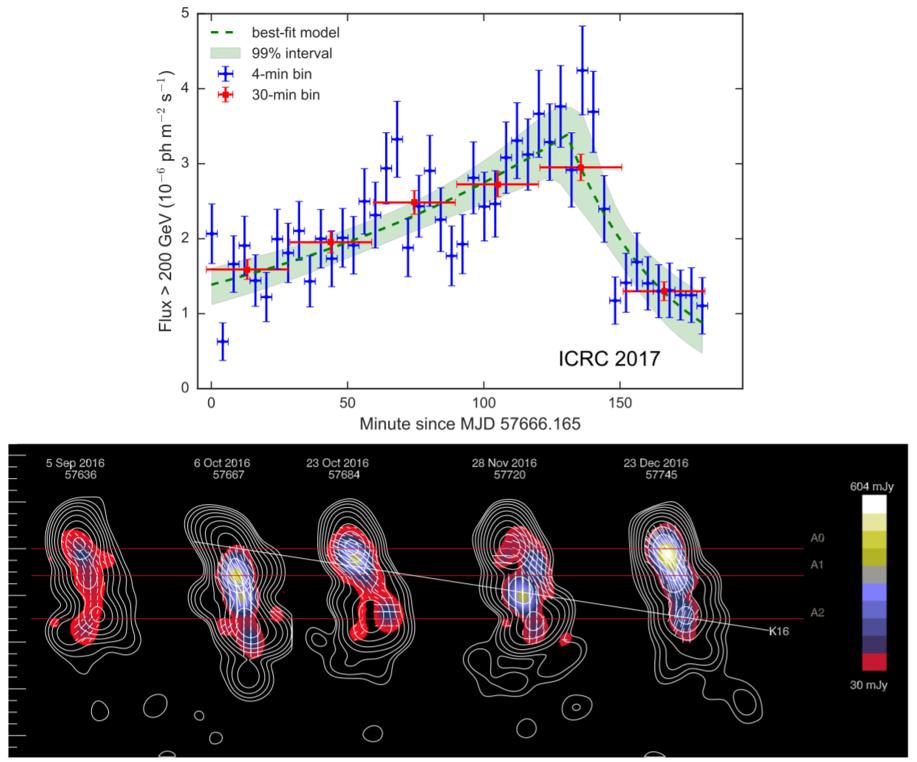}
\caption{The VHE gamma-ray light curve of BL Lac for energies higher than 200 GeV on October 5th (top). The blue dots show the light curve in 4-minute bins, and the red squares show it in 30-minute bins. The green dashed line and shaded region show the best-fit model and its 99$\%$ confidence interval. The lower panel shows intensity images of BL Lac from VLBA observations at 43 GHz from one month before the VHE gamma-ray flare to more than two months after the flare. The total intensity is shown in contours and linearly polarized intensity is shown in color. Red horizontal lines indicate the mean locations of the three quasi-stationary components, while the white line shows the motion of moving knot K16, which may be related to the gamma-ray flaring activity. } 
\label{Fig:BLLac_flare}
\end{figure}

Two flaring episodes were reported in BL Lac--one in 2015 reported by MAGIC and the other in 2016 reported by VERITAS. The outburst in 2016 was the stronger one and was triggered by VERITAS during regular monitoring~\cite{Feng648_GA83}. The maximum flux measured during the flare reached up to 180$\%$ of the Crab Nebula flux. As shown in Figure~\ref{Fig:BLLac_flare}, the flux was high enough to measure a lightcurve with 4-minute bins. The flare showed a smooth rising with a rise time of 140 minutes and shorter falling time of 36 minutes. During the VHE gamma-ray outburst, \textit{Fermi}-LAT also saw an elevated flux. But, the scale of flux change observed in \textit{Fermi}-LAT is on the order of day scale. Variability in the X-ray and optical fluxes, as well as changes in the optical and radio polarizations, were observed during the flaring. As shown in Figure~\ref{Fig:BLLac_flare}, a superluminal knot that may correspond to the VHE gamma-ray flaring was identified in the VLBA observations at 43 GHz. The detailed broadband SED, variability time scale measurements during the flare, and the polarization information can help us understand the mechanisms behind the VHE gamma-ray flare.

Another interesting flare reported during the conference was a flare of the radio galaxy NGC 1275 in 2016 reported by MAGIC~\cite{Glawion622}. The observations carried out over several months between 2016 and 2017 showed night-to-night flux variations, with the brightest flux reaching 1.75 times the flux of the Crab Nebula.


\section{Upgrades of existing experiments and future experiments}
Since the first detection of VHE gamma-ray emission from the Crab Nebula, VHE gamma-ray astronomy has progressed very rapidly together with our understanding of high-energy particles in our Universe. However, as we learn more about various sources and their natures, the limitations of current instruments become clearer. For example, still the largest portion of Galactic sources fall into the unidentified category, and a large number of these unidentified sources are located in complicated regions that make them difficult to identify firmly. Better sensitivity and angular resolution will help the identification. If the sensitivity of current IACTs improves by a factor of ten, then the Galactic plane survey pursued by H.E.S.S. over 3,000 hours will be reduced to 300 hours. Moreover, we may be able to find different stellar objects that can accelerate particles to the VHE band. The techniques of observing the air showers used to study the VHE gamma rays also have not been completely exploited. Achieving ten times better sensitivity with improved performance is a challenging but achievable goal for IACTs.

Here, I will summarize the planned upgrades of existing experiments and future experiments discussed during the conference.

\begin{figure}[t!]
  \centering
  \includegraphics[scale=0.4]{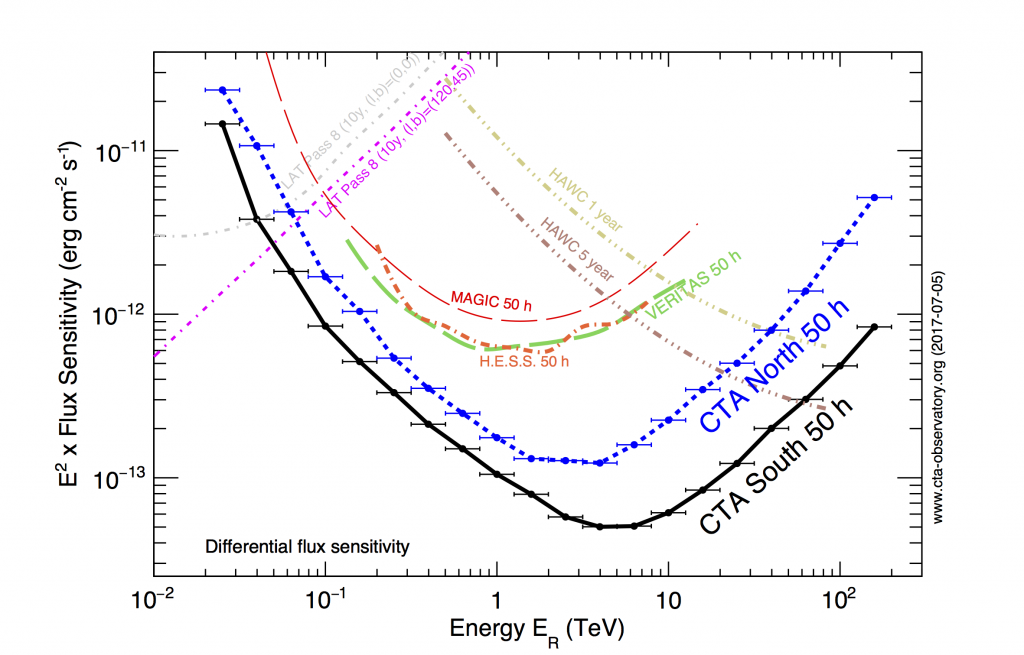}
\caption{Differential sensitivity of CTA to detect gamma rays with five standard deviations. Additional criteria applied are to require at least ten detected gamma rays per energy bin and a signal/background ratio of at least 20~\cite{CTA_web}.} 
\label{Fig:CTA_Sensitivity}
\end{figure}
\subsection{Future IACTs}
The Cherenkov Telescope Array (CTA)~\cite{Ong1071} is designed to be the next-generation major IACT observatory with up to an order of magnitude improved sensitivity compared to the current major IACTs arrays--H.E.S.S., MAGIC, and VERITAS. The performance goals of CTA include a good sensitivity, a wide energy range, full sky coverage, a wide FoV, excellent energy and angular resolution, and a rapid response time. All of the performance goals were set based on scientific motivations. The high sensitivity will impact on all science topics. Full-sky coverage will allow the characterization of the entire VHE gamma-ray sky. A wide FoV will enable rapid surveys and better studies for extended sources. Better angular resolution will allow detailed morphology studies, and better energy resolution will enable better spectrum measurements. The rapid response will enable quick follow-up observations for transient sources. To achieve the performance goals, CTA will deploy large arrays of telescopes to cover an area that is significantly larger than the Cherenkov light pool. To achieve a wide energy range, CTA will have a graded array of telescopes of different sizes--large-sized telescopes (LSTs) to extend energy measurements down to 20 GeV, medium-sized telescopes (MSTs) to cover the energy range from 100 GeV to 10 TeV, and small-sized telescopes (SSTs) to extend the measurement with good sensitivity up to > 300 TeV. The status of prototype studies presented during the conference can be found in \cite{Teshima_GA202} for the LSTs, \cite{Vassiliev838} for the MSTs, and \cite{Samarai800,Maccarone855,Helene822} for the SSTs. CTA will be built at two sites to cover the full sky. The southern site will be located in Chile with an array of four LSTs, 25 MSTs, and 70 SSTs. The northern site will be located on La Palma with an array of four LSTs and 15 MSTs. The expected sensitivity of both sites is shown in Figure~\ref{Fig:CTA_Sensitivity}. Extensive work has been carried out to prototype the hardware and software for all three telescope types, and was presented during the conference as well as the expected performance of CTA for each scientific case. CTA is planned to start construction in 2018. The status of prototype studies, calibration, simulation, analysis, and scientific studies were presented during the conference. 

The Monitoring at TeV energies (M@TE) project aims to achieve continuous monitoring of the VHE gamma-ray sky by installing FACT-like telescopes around the globe. This will close temporal gaps and compile light curves with homogeneous sensitivity~\cite{Alfaro776}. One of its goals is to refurbish a previous generation IACT, the HEGRA telescope, by upgrading all the electronics and changing the mirror components with a design that is similar to FACT. M@TE is planned to be installed at the Observatorio Astron\'{o}mico Nacional de San Pedro M\'{a}tir ($30^{\circ}54\arcmin19\arcsec$ N, $115^{\circ} 30\arcmin 04\arcsec$ W, 2200m asl). Fabrication and testing of the main components are underway.

\subsection{Future air shower arrays}
\begin{figure}[h!]
  \centering
  \includegraphics[scale=0.5]{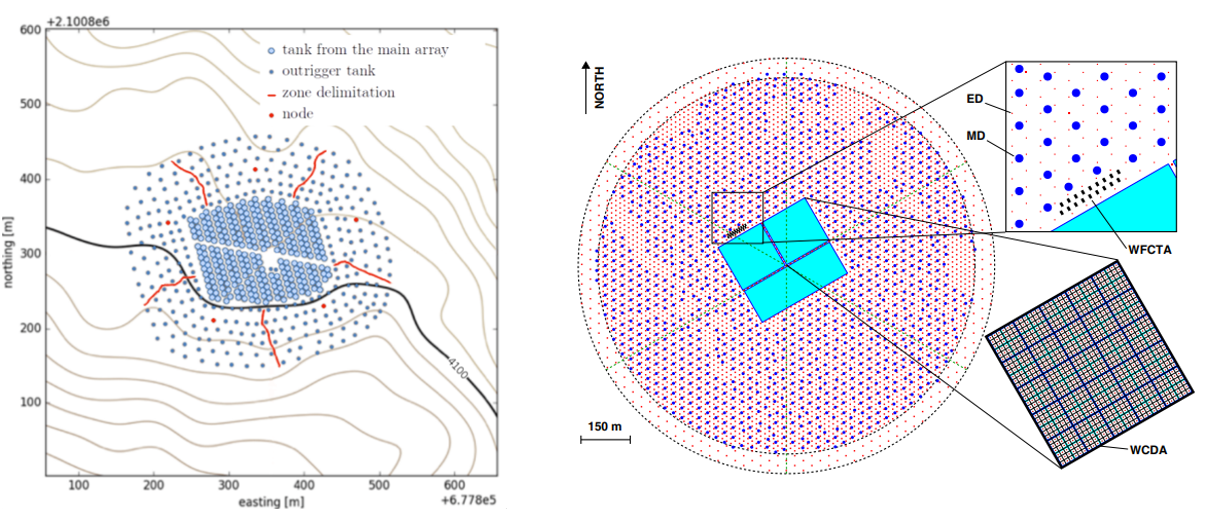}
\caption{The layout of outriggers for the HAWC upgrade (left)~\cite{Joshi806_GA149}, and the layout of LHAASO, a future air shower array (right)~\cite{Lv455}.} 
\label{Fig:Future_AirShowerArray}
\end{figure}

HAWC is updating their array by deploying outriggers~\cite{Joshi806_GA149}, a large array of small water Cherenkov detectors surrounding the main HAWC array. Air showers from gamma rays with energies of tens of TeV will activate all of the detectors in the main HAWC array. Particles from these high-energy gamma rays may not be well contained in the array, and it is hard to reconstruct the core locations of these particles. The outrigger array for HAWC is designed to provide a good core location measurement for high-energy air showers by increasing the size of the array. This will enhance the high-energy sensitivity of HAWC. The outrigger array consists of 350 cylindrical tanks of diameter 1.55 m. As shown in the left side panel of Figure~\ref{Fig:Future_AirShowerArray}, HAWC with the outriggers will cover an area about 4--5 times larger than the main array alone. Currently, the deployment of the outrigger array is taking place, and the first few outriggers are already taking data on site. The full outrigger array is planned to be completed by the beginning of 2018.

The Large High Altitude Air Shower Observatory (LHAASO) is a future hybrid extensive air shower array to be constructed at Mount Haizi ($100.01^{\circ}$ E, $29.35^{\circ}$ N, 4410m a.s.l.) in Sichuan province, China. The array is designed to measure cosmic rays in the energy range between 1 TeV and 1 EeV and gamma rays in the energy range between 100 GeV and 1 PeV. It consists of an area of 1 $\textnormal{km}^{2}$ array (KM2A), a water Cherenkov detector array (WCDA), and a wide-FoV imaging Cherenkov telescope array (WFCTA)~\cite{Liu424}. KM2A consists of about 5,000 plastic scintillators to detect electromagnetic particles and about 1,000 water Cherenkov detectors under 2.5 m soil as muon detectors. Located at the center of the array and covering a total area of 78,000 $m^{2}$, the WCDA will consist of three large pools divided into $\sim$3,000 optically isolated cells. The WFCTA will include 12 telescopes with FoV $14^{\circ} \times 16^{\circ}$ that will operate to measure both Cherenkov radiation and the fluorescence light to study the wide energy range of cosmic-ray showers. A partial detector of the WCDA will start operating from late 2018, aiming to have the full array of WCDA operating by the end of 2020~\cite{Chen832}. The layout of the experiment can be found on the right side panel of Figure~\ref{Fig:Future_AirShowerArray}. 

The present status and plans for a future expansion were reported for the Tunka Advanced Instrument for cosmic ray physics and Gamma Astronomy (TAIGA)~\cite{Budnev768}. TAIGA is located in the Tunka valley in Siberia ($51^{\circ}48\arcmin47.5\arcsec$ N, $103^{\circ}04\arcmin16.3\arcsec$ E, 675 m asl). TAIGA is designed to study gamma rays and cosmic rays in the energy range from 10 TeV to 1 EeV. Currently, TAIGA consists of an array of 28 air Cherenkov timing stations, HiSCORE (High Sensitivity Cosmic Origin Explorer). By the end of 2019, TAIGA is aiming to expand the array to include 100-120 stations in an area of $\sim$1 km$^{2}$, three IACTs, and 200 m$^{2}$ of muon detectors. The expected integral sensitivity of TAIGA 1 $km^{2}$ with 300 hours of exposure above 100 TeV will be about $2.5\times10^{-13}$ TeV cm$^{-2}$ sec$^{-1}$.


\section{Summary}
After more than ten years of observations with the current generation's IACTs, we are still finding large numbers of new VHE gamma-ray sources in the sky. Since the last ICRC, the number of VHE gamma-ray sources has increased by 36, and the total number of VHE gamma-ray sources counts 198. About half of the new sources were detected by HAWC as they accumulated more than two years of data with a full detector configuration. HAWC's new sources present different characteristics than sources that were detected by IACTs. Because HAWC has a large FoV and covers 2/3 of the sky, HAWC can provide a population of very extended sources including local sources that are closer than 1 kpc to us. These are particularly interesting in connection to the leptonic cosmic-ray fluxes we measure at Earth. For IACTs, we achieve one more milestone for the technique by a firm detection of the extension of the Crab Nebula, which was treated as a point-like source up to now. The first binary from outside of our Galaxy has been detected, and new flaring episodes observed in wide electromagnetic bands allow us detect new VHE gamma-ray sources and study the extreme environments around AGNs. Detailed studies of known sources reveal the particle acceleration and interaction in different environments. Currently we know more about the VHE accelerators than ever before with the best sky coverage in history. There will be another leap in our understanding of the VHE gamma-ray universe once CTA comes online. We will learn about the in-depth details of sources we are studying now. More importantly, CTA will reveal more new sources, which likely will include at least one new class of accelerators we have not yet detected. The construction of CTA will start in 2018, so the future is not very far away. 


\end{document}